\documentclass[
superscriptaddress,
preprint,
notitlepage,
amsmath, amssymb,
aps,
prab,
floatfix,
]{revtex4-2}
\usepackage{graphicx,hyperref}

\begin{document}

\preprint{SLAC-PUB-251125}

\title{Analytical Approximations for Beamstrahlung at
                  Very High Energy Electron-Positron Colliders}

\author{Dongxing He}
\email{jhe5718@umd.edu}
\affiliation{University of Maryland, College Park, College Park, Maryland 20742, USA}
\author{Arianna Formenti}
\thanks{Work supported by the US Department of Energy, contract HEP-ERCAP0031191.}
\affiliation{Lawrence Berkeley National Laboratory, Berkeley, California 94720, USA}
\author{Spencer Gessner}
\thanks{Work supported by the US Department of Energy, contract DE--AC02--76SF00515.}
\affiliation{SLAC National Accelerator Laboratory, Menlo Park, California 94025, USA}
\author{Michael Peskin}
\email{mpeskin@slac.stanford.edu}
\affiliation{SLAC National Accelerator Laboratory, Menlo Park, California 94025, USA}
\date{\today}

\begin{abstract}
Among the many effects that occur in beam-beam electron-positron collisions at TeV energies, emission of hard synchrotron radiation, or beamstrahlung, has special importance. Beamstrahlung determines the energy spectrum of the most energetic electrons, positrons, and photons and supplies the initial condition for the calculation of all other QED processes. In this paper, we show that the description of beamstrahlung simplifies in the limit of large quantum parameter $\Upsilon$, which is realized in 10~TeV collider designs. The beamstrahlung spectra for electrons and photons are given in terms of universal functions. We supply approximations to these functions that will be useful for more general studies of the beam-beam interaction at very high energies.
\end{abstract} 

\maketitle
\newpage

\section{Introduction}

Among the technologies that are considered today for the step to the next energy frontier -- parton collisions at 10~TeV energies -- wakefield-based linear colliders are one of the most promising \cite{Esarey:2009zz, Lindstrom:2025yng, Lu:2022}. Plasma acceleration experiments have demonstrated accelerating gradients greater than GVm$^{-1}$, which would give access to the 10~TeV scale with a relatively compact machine design. A design study for a practical realization of a 10~TeV wakefield collider is now underway~\cite{Gessner:2025acq}.

Still, many difficulties with this technology need to be overcome.  A linear collider operates by colliding trains of electron and positron bunches, with $10^9$-$10^{10}$ particles per bunch. An important difficulty is the fact that collisions of these particle bunches entail very strong QED interactions among the colliding particles. This leads to the production of energetic photons and $e^+e^-$ pairs and the associated broadening of the colliding bunch energy spectra. This is the result of the small size and high density of the colliding $e^+e^-$ bunches, required to produce target luminosities in excess of $10^{35}$ cm$^{-2}$s$^{-1}$.   Among several contributing effects, the most important one is beamstrahlung, or hard synchrotron radiation due to the intense electric fields that the particles of one bunch encounter while passing through the other~\cite{Augustin:1978ah, Himel:1985bm, Blankenbecler:1987rg, Jacob:1987mg, Jacob:1987ua, Bell:1987rw, Bell:1987ve}. Beamstrahlung is a major source of energy loss for each bunch and a major contributor to beam-related background processes. It also sets up the initial conditions for calculations of other QED processes, such as $e^+e^-$ pair production, in the bunch-bunch interaction. In this paper, we will concentrate on the analysis and modeling of beamstrahlung.  Some comments on other contributing effects are given in the conclusions.

To learn how to mitigate beamstrahlung, we must first understand it better. The study of beamstrahlung began in the 1980s, in preparation for the Stanford Linear Collider (SLC). It was recognized that the design parameters of this collider brought one into the regime in which quantum effects in synchrotron radiation would become important~\cite{Himel:1985bm}. These quantum effects are characterized by the parameter
\begin{eqnarray}
  \Upsilon = {\gamma^2\over m_e\rho}  =  {2\over 3}{\omega_c\over E_b} \ ,
  \label{Upsdef}
\end{eqnarray}
where $E_b = \gamma m_e$ is the beam energy, $\rho$ is the radius of curvature due to the fields of the opposite bunch, and $\omega_c$ is the classical critical energy for synchrotron radiation. Note that in the collider physics literature, we use $\Upsilon$ to describe this parameter, while in the Strong Field QED literature \cite{Gonoskov:2022, FEDOTOV20231}, $\chi$ is used for this quantity. The criterion for the quantum regime is $\Upsilon > 1$.   In linear collider designs, $\Upsilon$ is of order 1 for $e^+e^-$ Higgs with CM energies below 1~TeV.   At 10~TeV, $\Upsilon$ can be of the order of $10^3$~\cite{Barklow:2023}.

Around 1990, Yokoya and Chen developed a theory of multi-photon beamstrahlung emission and validated this against simulations~\cite{Yokoya:1989jb, Yokoya:1991qz}. The Yokoya-Chen analytic theory and subsequent simulation codes, GUINEA-PIG by Schulte~\cite{Schulte:1999tx} and CAIN by Yokoya and collaborators~\cite{Chen:1994jt}, led to an understanding of beamstrahlung sufficient to design $e^+e^-$ Higgs factories at center-of-mass energies from 90 to 500 GeV.  At these energies, with flat beam designs, the electrons lose less than 5\% of their energy to beamstrahlung, an amount comparable to the loss from initial state radiation \cite{Frixione:2022ofv}.

Today, we are motivated by the 2023 P5 Report~\cite{P5_report} to consider $e^+e^-$ collisions with 10~TeV center-of-mass energy. Preliminary designs for wakefield colliders have bunch-bunch collisions deep in the quantum regime for beamstrahlung, with a substantial fraction of the total energy in electrons and positrons lost to beamstrahlung radiation~\cite{Barklow:2023}. For these colliders, a better understanding of beamstrahlung is needed to understand the discovery potential of the collider.  Some steps toward this goal have been taken in \cite{DelGaudio:2019, Zhang:2023, Yakimenko:2018kih, Zhang_2025}, using more advanced simulations to probe the more strongly quantum regime at higher energies. 


In this paper, we develop a theory of beamstrahlung applicable to colliders with CM energies of order 10~TeV. We begin with a discussion of scaling laws for beamstrahlung in Section~\ref{sec:scaling}.   We then analyze beamstrahlung in three stages.   In Section~\ref{sec:simulations} we describe simulations of the bunch-bunch interaction for 10 TeV wakefield collider parameters using the GUINEA-PIG code. In Section \ref{sec:YC}, we describe the Yokoya-Chen model and compare it to the simulation results.  
In Sections~\ref{sec:toAYC}-\ref{sec:AYC}, we show that the Yokoya-Chen model simplifies and reaches an asymptote in the limit of large $\Upsilon$. This gives a description of beamstrahlung that depends only on $N_\gamma$, the expected number of beamstrahlung photons produced from the highest energy electrons and positrons, with energy distributions of the particles in the bunch described by universal functions. In Section~\ref{sec:slices}, we work out the consequences of this asymptotic theory in a simple model of the time evolution of colliding bunches. We show that this theory, used according to our model, gives a good description of the simulation data for the luminosity distribution and a more general set of observables. In Section~\ref{sec:timeforlumi}, we use the asymptotic theory to describe the time evolution of the production of luminosity, both for $e^+e^-$ and for $e\gamma$ and $\gamma\gamma$ collisions. Section~\ref{sec:conclusions} give some conclusions of this analysis. We use natural units with $c=1$ throughout the paper.

\section{Scaling laws for beamstrahlung}
\label{sec:scaling}

Our updated analytic theory of beamstrahlung should provide a dynamic description of the radiative process during a single bunch-bunch collision, and it should describe changes to figures-of-merit, such as $N_\gamma$, as collider parameters are changed. We start by considering scaling laws that apply within a single bunch-bunch collision. Here we fix the initial beam energy $\gamma_b$, while allowing the energy of individual particles $\gamma$ to change throughout the collision. All other beam parameters, including the number of particles per bunch $N$, the normalized emittances $\epsilon_a$, the beta functions $\beta_a$, with $a = x, y$, and the bunch length $\sigma_z$, are held constant. 





A relativistic particle bending on an arc with radius of curvature $\rho$ will have the classical rate of emission of synchrotron photons 
\begin{eqnarray}
\nu_{cl} =  {5\over 2\sqrt{3}} \, \alpha \, {\gamma\over \rho}.
\label{nucldef} \end{eqnarray}
The bending radius $\rho$ is given by 
\begin{eqnarray}
\rho =  \frac{m\gamma}{2eB},
\label{rhodef} \end{eqnarray}
where $B$ is the magnetic field strength of the opposing beam. Since, for fixed forces, $\rho\propto \gamma$, $\nu_{cl}$ is independent of particle energy within a given bunch collision, even as the particles radiate beamstrahlung photons. 

However, the energy of the radiated photons grows rapidly if we raise the energy of the collisions. The classical critical energy is given by
\begin{eqnarray}
\omega_c = {3\over 2} {\gamma^3\over \rho}.
\end{eqnarray}
This scales as $\gamma^2$ because $\rho\propto \gamma$. Eventually, $\omega_c$ becomes comparable to the particle beam energy.  At this point, the classical formulae must be modified to take into account that an electron can not radiate a photon with energy greater than its own \cite{Ternov_1995}. For a collider design, this transition from the classical to the quantum formulae for synchrotron radiation is described by the collider parameter
\begin{eqnarray}
\Upsilon = {2\over 3} {\omega_c\over E_b},
\end{eqnarray}
where $E_b$ is the nominal beam energy.

In a collision of Gaussian bunches, the average value of $\Upsilon$ at the first emission is given by \cite{Chen:1992}
\begin{eqnarray}
 \Upsilon_{avg} = {5 N r_e^2 \gamma_b\over 6 \alpha \sigma_z (\sigma_x
  + \sigma_y)}, 
\label{Upsilonavg}
\end{eqnarray}
where $r_e$ is the classical electron radius. The bunch sizes depend on the beta functions  and the normalized emittances according to
\begin{eqnarray}
    \sigma_a = \sqrt{\beta_a \epsilon_a/\gamma_b} \propto
    1/E_b^{1/2},
    \end{eqnarray}
so that
    \begin{eqnarray}
    \Upsilon \propto  E_b^{3/2}.
\end{eqnarray}
The bunch sizes during a collision are also affected by the mutual attraction of electrons and positrons in the colliding bunches. This is parametrized by the disruption parameters  $D_a$ \cite{Chen_Disruption:88}, given by
\begin{eqnarray}
    D_a =  {2 N r_e \sigma_z\over \gamma_b  \sigma_a (\sigma_x + \sigma_y)}.
\end{eqnarray}
We consider now the scaling of collider energy while holding all other parameters fixed.  In this case, the $D_a$ are constant as $E_b$ is increased, so that the distortions of the bunch shapes remain the same as $E_b$ is varied. Then we expect the luminosity, beamstrahlung, and disruption parameters to scale with beam energy as
\begin{eqnarray}
 {\cal L}&\propto& E_b^1, \nonumber \\
 \Upsilon &\propto& E_b^{3/2}, \nonumber \\ 
 D_a &\propto& E_b^0. 
  \label{accscaling}
\end{eqnarray}
In the next section, we will describe two sets of simulations of the bunch-bunch interaction for multi-TeV energy colliders, with flat and round bunch profiles. In both cases, $\epsilon_a$ and $\beta_a$ are held fixed as the CM energy is increased.

At high values of $\Upsilon$, the rate of radiation by a single particle decreases \cite{Ternov_1995}.  We will see in Section~\ref{sec:toAYC} that the radiation rate described by the quantum formulae varies with beam energy $\nu(E_b)$ as 
\begin{eqnarray}
    \nu(E_b)  \propto \nu_{cl} \Upsilon^{-1/3}.    \label{nuEb}
\end{eqnarray}
We note that the general form of the beam field of a Gaussian bunch is given by \cite{Bassetti:1980}
\begin{eqnarray}
    B\propto \frac{1-e^{-a^2/2\sigma_a^2}}{a}.
\end{eqnarray}
Therefore, $\rho\propto E_b^{-1/2}$ as the accelerator energy is increased. Then, according to equation \ref{nucldef}, $\nu_{cl} \propto E_b^{1/2}$ for changing \emph{beam} energy. With this scaling of $\nu_{cl}$ and $\Upsilon$ scaling as in equation \ref{accscaling}, we then find that $\nu(E_b)$ is independent of the collider energy. The nominal number of photons radiated by an electron at the beam energy,
\begin{eqnarray}
N_\gamma = \nu(E_b) \cdot \sqrt{3} \sigma_z ,
\label{Ngamma}
\end{eqnarray}
is also independent of beam energy.

These scaling laws tell us that, as the accelerator energy is raised with all other parameters held constant, the amount of beamstrahlung radiation will be very similar. The focusing of the bunches due to disruption and the total number of high-energy photons emitted will be unchanged as the collider CM energy is increased. The only differences in the beamstrahlung rate will come from the increase of $\Upsilon$.

\section{Simulations}
\label{sec:simulations}

\begin{figure}
\begin{center}
\includegraphics[width=0.9\hsize]{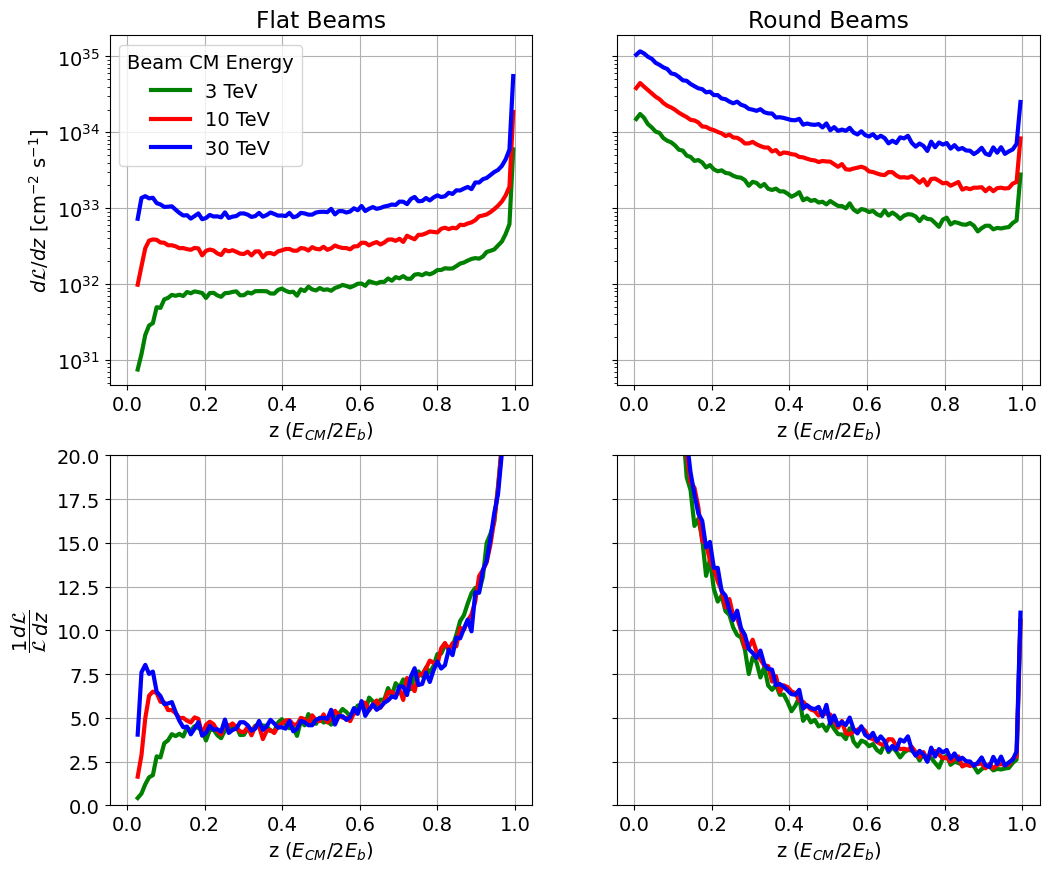} 
 
\end{center}
\caption{Luminosity distributions $d{\cal L}/dz$ from simulation, comparing designs at 3, 10, and 30~TeV with beam parameters from Table~\ref{tab:params}. Top: Un-normalized distributions plotted on a log scale. Bottom: Normalized distributions plotted on a linear scale. Left: flat beam parameters; Right: round beam parameters.}
\label{fig:showlumi}
\end{figure}

To establish a baseline, we carried out simulations of $e^+e^-$ bunch-bunch collisions for two series of accelerator designs for colliders at 6 different CM energies ranging from 3~TeV to 30~TeV. Within each series, the parameters $\beta_a$, $\epsilon_a$, $N$, and $\sigma_z$ were held constant. The two series represented different aspect ratios for the colliding bunches: flat beams with $\sigma_y/\sigma_x = 0.027$ and round beams with $\sigma_y/\sigma_x = 1$.

For each case, the actual $e^+e^-$ collisions take place between particles with, generally, lower energy than the nominal beam energy $E_b$ and lower CM energy than the nominal $2E_b$. We can describe individual $e^+e^-$ particle collisions using the $e^+$ or $e^-$ normalized energy, $x = E_{+, -}/E_b$, the normalized CM energy, $z = E_{CM}/(2E_b)$, and the CM rapidity, $y = \frac{1}{2}\log\left(E_-/E_+\right)$, and similarly for $e^\pm \gamma $ and $\gamma\gamma $ collisions with in the bunch crossing. 

The accelerator parameters for these simulations are given in Table~\ref{tab:params}. The simulations were performed using the GUINEA-PIG particle-in-cell (PIC) simulation code~\cite{Schulte:1999tx}. More details are supplied here~\cite{peskin_2025_17716686}.
These simulations are performed with incoherent and coherent QED processes turned off, except for beamstrahlung emission. The computed luminosities are quoted in Table~\ref{tab:outparams}. We also show the geometric luminosity ${\cal L}_{geo}$ and quote ${\cal L}_{20}$, the luminosity for collisions within 20\% of the nominal CM energy ($z > 80\%$), which is a proxy for the luminosity usable for discovery physics~\cite{Barklow:2023}.



Figure~\ref{fig:showlumi} shows the absolute luminosity distribution $d {\cal L}/dz$  for the cases of 3, 10, and 30 TeV collider center-of-mass energies. Each plot compares the flat and round beam cases. Remarkably, there is an asymptotic curve realized for large $\Upsilon$ that is reached rapidly as the CM energy increases above 1~TeV. We will explain the origin of this asymptote in the following sections.



 \begin{table}[b]
\caption{\label{tab:params}Fixed beam parameters for the simulations described in this paper.}
\begin{ruledtabular}
    
    \begin{tabular}{lcc}
    \hline
    &   \textbf{Flat beams}   &  \textbf{Round beams} \\ \hline
    $\epsilon_x$ (mm rad) &  0.66   & 0.1\\
    $\epsilon_y$ (mm rad) &  0.02   & 0.1\\
    $\beta^*_x$ (mm) &  5.0  & 0.15\\
    $\beta^*_y$ (mm) &  0.1 & 0.15\\
    $\sigma_z$ ($\mu$m) & 5  & 5\\
    $N$ ($\times 10^9$) &  5   & 5\\

    $N_\gamma$   &   1.5          &   5.7      \\
    
\hline
   \end{tabular}
   \end{ruledtabular}
\end{table}


\begin{table}[b]
\caption{\label{tab:outparams}Luminosities from simulations of different energy colliders: ${\cal L}_{geo}$ is the nominal or geometrical luminosity;  ${\cal L}_{sim}$ is the total luminosity from simulation.  ${\cal L}_{20}$ is the luminosity for particle collisions with  $z = E_{CM}(e^+e^-)/2E_b > 80\%$. Luminosities are quoted in units of cm$^{-2}$sec$^{-1}$, using the repetition rates $f$ given in the previous table. $\Upsilon_{\text{sim}}$ is the average value of $\Upsilon$ outputted from the GUINEA-PIG simulation, given by $5/12 \cdot \Upsilon_{\text{max}}.$}
\begin{ruledtabular}
\begin{tabular}{lccc}
\hline
\textbf{CM Energy (TeV)} & \textbf{3} & \textbf{10} & \textbf{30} \\ 
\hline
\multicolumn{4}{l}{Flat beams:} \\
${\cal L}_{geo}$ ($\times 10^{35}$) & 0.94 & 3.13 & 9.40 \\
${\cal L}_{sim}$ ($\times 10^{35}$) & 1.71 & 5.78 & 17.4 \\
${\cal L}_{20}$ ($\times 10^{35}$)  & 1.05 & 3.34 & 9.91 \\
$\Upsilon_{\text{sim}}$ & 87.2 & 544. & 2844. \\
\multicolumn{4}{l}{Round beams:} \\
${\cal L}_{geo}$ ($\times 10^{35}$) & 3.02 & 10.1 & 30.2 \\
${\cal L}_{sim}$ ($\times 10^{35}$) & 25.7 & 77.1 & 222. \\
${\cal L}_{20}$ ($\times 10^{35}$)  & 1.38 & 4.36 & 13.5 \\
$\Upsilon_{\text{sim}}$ & 1166. & 6705. & 34387. \\
\hline
\end{tabular}
\end{ruledtabular}

\end{table}


\section{The Yokoya-Chen description of beamstrahlung}
\label{sec:YC}

Yokoya and Chen suggested an evolution equation for the energy distribution of electrons and positrons resulting from beamstrahlung \cite{Yokoya:1989jb, Chen:1992}.  Their starting point was the Sokolov-Ternov formula for the probability of a single electron with an energy in the quantum regime to emit synchrotron radiation \cite{ST1, ST2}.  To write this formula, it is convenient to define some parameters of synchrotron radiation as a function of the energy of the radiating particles. 

Recall from Section~\ref{sec:scaling} that, for a particle at a given energy $E$, the classical critical energy increases as $E^2$. It is convenient to define a parameter $\kappa = E^2/\omega_c$, which is independent of energy and contains the dependence on the EM field. Then define
\begin{eqnarray}
\xi(E) = { \omega_c\over E} = {E\over  \kappa}
\end{eqnarray}
to represent the energy of a particle in the bunch-bunch collision for the purpose of computing its rate of synchrotron radiation. The value of $\xi$ at the beam energy is related to the collider parameter $\Upsilon$ by 
\begin{eqnarray}
\Upsilon = {2\over 3} \xi(E_b) \ .
\end{eqnarray}

The radiation rate in the quantum regime for an electron of energy $\bar E$ to emit a synchrotron photon and transition to an energy $E$ is then given by the formula~\cite{ST1,ST2}
\begin{eqnarray}
           F(\bar E, E) = {3\over 5\pi} {\kappa  \nu_{cl}\over \bar E
             E}\biggl[
           {1\over 1 + \bar \xi y }\int^\infty_y K_{5/3}(y') dy' +
             {(\bar \xi y)^2\over (1 + \bar \xi y)^2}K_{2/3}(y)
             \biggr]    ,\label{STeq}
             \end{eqnarray}
where $\xi = E/\kappa$, $\bar \xi = \bar E/ \kappa$, $y = \kappa (1/E - 1/\bar E)$, and $K_\nu(y)$ is the modified Bessel function \cite{NIST:DLMF}. The emission rate $\nu(\bar E)$ for an electron of energy $\bar E$ to radiate to any lower energy is equal to the integral of the right-hand side of (\ref{STeq}) over $E \in (0, \bar{E})$.
             
Assume that the field through which the electron travels is approximately uniform and that successive emissions are uncorrelated. The distribution of energies in the electron bunch can then be described by a distribution function $h_e(x,t)$, with $x = E/E_b$.  This distribution function satisfies the initial condition
\begin{eqnarray}
            h_e(x,t=  0 ) = \delta(x-1)     \label{yinit}
\end{eqnarray}
and satisfies the normalization condition
 \begin{eqnarray}
            \int^1_0 dx \ h_e(x,t) = 1     \label{ynorm}
\end{eqnarray}
for all times $t$. Under the assumptions above,  the formula (\ref{STeq}) implies that $h_e(x,t)$ satisfies the differential
equation
            \begin{eqnarray}
            {\partial\over \partial t} h_e(x,t)  &=& - \nu(x) h_e(x,t) \nonumber   \\
                  && \hskip -0.4in  + \int_x^1 d\bar E \  {3\over 5\pi}  {\kappa
                    \nu_{cl}\over E \bar E}\ \biggl[
           {1\over 1 + \bar \xi y }\int^\infty_y K_{5/3}(y') dy' +
             {(\bar \xi y)^2\over (1 + \bar \xi y)^2}K_{2/3}(y)
             \biggr] \ h_e(\bar x, t)
             \label{ycdeq}
             \end{eqnarray}
where $x = E/E_b$ and $\bar x = \bar E/E_b$. The first term on the r.h.s of (\ref{ycdeq}) is a sink term removing high-energy electrons from the distribution, and the second term is a source that reintroduces the electrons to the distribution at lower energy after radiation. Simplify this using $\kappa/E = \xi_b^{-1} /x = 2/( 3\Upsilon x)$ and 
\begin{eqnarray}
             \bar\xi y = {\bar x (\bar x - x)\over x\bar x} = {\bar
               x\over x} - 1 \ .
\end{eqnarray}
This gives
        \begin{eqnarray}
            {\partial\over \partial t} h_e(x,t)  &=& - \nu(x) h_e(x,t) \nonumber   \\
                  && \hskip -0.4in  +  {2   \nu_{cl}\over 5\pi}
                  \Upsilon^{-1}\int_x^1{ d\bar x\over x \bar x}  \ 
         \biggl[
           {x\over \bar x }\int^\infty_y K_{5/3}(y') dy' +
             {(\bar x - x)^2\over \bar x^2}K_{2/3}(y)
             \biggr] \ h_e(\bar x, t) \ ,     \label{YCeq}
             \end{eqnarray}
which is the Yokoya-Chen equation for multi-photon beamstrahlung~\cite{Yokoya:1989jb, Yokoya:1991qz}.

\section{The Yokoya-Chen model at large $\Upsilon$}
\label{sec:toAYC}           

The data in Fig.~\ref{fig:showlumi} shows that for large values of $\Upsilon$, the shape of the luminosity spectrum remains fixed when scaling the CM energy while holding $\beta_a,~\epsilon_a,~N$, and $\sigma_z$ constant. This suggests that we should study the Yokoya-Chen equation for very large values of $\Upsilon$.  Notice that 
\begin{eqnarray}
y = \frac{1}{\xi_b}\left(\frac{1}{x} - \frac{1}{\bar{x}}\right) = \frac{2}{3\Upsilon}\left(\frac{1}{x} - \frac{1}{\bar{x}}\right)   
\end{eqnarray}
in (\ref{YCeq}), and thus becomes small in this limit for all but the most energetic radiation events. This motivates taking the limit of the Bessel functions for small argument. Substituting,
\begin{eqnarray}
    K_{2/3}(y)& \to& {2^{2/3} \over 2}\mbox{\small $\Gamma({2\over 3})$} 
    y^{-2/3} = { 3^{2/3}\over 2}\Upsilon^{2/3}
    \mbox{\small $\Gamma({2\over 3})$}\left( { x \bar x\over \bar x - x}\right)^{2/3} \nonumber   \\
    \int^1_y dy' K_{5/3}(y') & \to& 2^{2/3}\mbox{\small $\Gamma({2\over 3})$} 
    y^{-2/3} =  3^{2/3}\Upsilon^{2/3}
    \mbox{\small $\Gamma({2\over 3})$} \left( { x \bar x\over \bar x - x}\right)^{2/3} \ ,
\end{eqnarray}
which gives
\begin{eqnarray}
    {\partial\over \partial t} h_e(x,t)  &=& - \nu(x) h_e(x,t) \nonumber   \\
    && \hskip -0.4in  +  {3^{2/3}  \Gamma({2\over 3})\nu_{cl}\over 10\pi}
    \Upsilon^{-1/3}\int_x^1 d\bar x\
    {({\bar x}^2 + x^2)/{\bar x}^2\over  x^{1/3} (\bar x - x)^{2/3} \bar x^{1/3}}
    \  h_e(\bar x, t) \ .    \label{YCinterm}
\end{eqnarray}
This shows explicitly the $\Upsilon^{-1/3}$ scaling of the photon emission rate suggested in (\ref{nuEb}).

We may simplify it further with some additional approximations that are valid for $x$, $\bar x$ near 1, corresponding to soft radiation events. For the first of these, replace in the numerator
\begin{eqnarray}
    ({\bar x}^2 + x^2)/{\bar x}^2 \to  2   \ .    \label{YCapprox}
\end{eqnarray}
To find $\nu(x)$, integrate the second term on the right-hand side over $x$.  We need the integral
\begin{eqnarray}
  \int^1_0 dx\ \int_x^1 d\bar x\
           {1\over  x^{1/3} (\bar x - x)^{2/3} \bar x^{1/3}}
           \    h_e(\bar x, t)  &=&  \int^1_0 d\bar x\ h_e(\bar
           x,t) \cdot \int^{\bar x}_0 dx    {1\over  x^{1/3} (\bar x -
             x)^{2/3} \bar x^{1/3}}\nonumber   \\
           &=&   \int^1_0 d\bar x\ h_e(\bar
           x,t) \cdot  {\Gamma({1\over 3}) \Gamma({2\over
               3})\over x^{1/3}}
\end{eqnarray}
Note that $ \Gamma({1\over 3}) \Gamma({2\over 3}) =  2\pi/\sqrt{3}$. Then, to ensure that our approximate YC equation conserves the normalization of $h_e(x,t)$, the total emission rate $\nu(E)$ should be
\begin{eqnarray}
  \nu(E) =  { 4\cdot 3^{1/6} \Gamma({2\over 3})\over 5 } \nu_{cl} \Upsilon^{-1/3} /x^{1/3}=
  1.301\  \nu_{cl} \Upsilon^{-1/3}/x^{1/3} \,
  \label{emmrate}
\end{eqnarray}
which agrees with equation 3.62 in reference~\cite{Yokoya:1991qz} at the 30\% level due to the approximation in (\ref{YCapprox}). 
The emission rate (\ref{emmrate}), at the beam energy ($x=1$), integrated through the bunch collision, is $N_\gamma$. 
In this approximation,
\begin{eqnarray}
     N_\gamma =  { 4\cdot 3^{1/6} \Gamma({2\over 3})\over 5 } \  \nu_{cl}
     \Upsilon^{-1/3} \cdot \sqrt{3} \sigma_z  \ . 
\end{eqnarray}

In realistic bunch-bunch collisions, the bunches have a Gaussian profile in all spatial coordinates. However, in the simulations, disruption and pinch effects pull the bunches into a narrower cylindrical region. 
For simplicity, we model the bunch-bunch collision as a collision between two cylindrical bunches, with sharp cutoffs at the front and back. We take the length of each cylinder to be the Gaussian expectation $\sqrt{3} \sigma_z$. To implement this, multiply (\ref{YCinterm}) by $\sqrt{3} \sigma_z$ and the rescale the time by defining
\begin{eqnarray}
    \tau = t \cdot N_\gamma/\sqrt{3} \sigma_z \ .
\end{eqnarray}
Then $\tau$ will run over the interval $[0, N_\gamma]$ during the bunch crossing.  This gives the final form of the asymptotic Yokoya-Chen equation:
\begin{eqnarray}
{\partial\over \partial \tau} h_e(x,\tau)  = -  h_e(x, \tau)/x^{1/3}
     +  {1\over B} \int_x^1 d\bar x\
       {1\over  x^{1/3} (\bar x - x)^{2/3} \bar x^{1/3}}
  \    h_e(\bar x, \tau) \ ,
  \label{AYC}
\end{eqnarray}
where $B = 2\pi/\sqrt{3}$. Every transition of an electron or positron gives rise to a photon.  In a pure beamstrahlung analysis, photons, once emitted, do not change their energy, and these photons accumulate in the process of the bunch collision.  Following the same logic as above, the energy distribution function for photons satisfies the equation
\begin{eqnarray}
 {\partial\over \partial \tau} h_g(x,\tau)  = 
                 + {1\over B} \int_x^1 d\bar x\
                   {1\over  x^{2/3} (\bar x - x)^{1/3} \bar x^{1/3}}
                   \    h_e(\bar x, \tau) \ ,
                   \label{AYCg}
                   \end{eqnarray}
                   with the initial condition
                   \begin{eqnarray}
                     h_g(x,\tau = 0) = 0 \ .
                     \label{gammainit}
\end{eqnarray}            
It can be shown that (\ref{AYC}) and (\ref{AYCg}) together imply
\begin{eqnarray}
   {\partial\over \partial \tau} \ \int_0^1 dx \ x \ ( h_e(x,\tau) +
   h_g(x,\tau))  = 0 \ ,
\end{eqnarray}
that is, conservation of the total energy throughout the collision. The equations (\ref{AYC}) and (\ref{AYCg}) give a closed Asymptotic Yokoya-Chen (AYC) model for the emission of beamstrahlung in a bunch-bunch collision.

This result is remarkable. Equation (\ref{AYC}) has no free parameters. It applies for all sufficiently large values of $\Upsilon$, and the value of $N_\gamma$ is encoded in the length of the evolution in $\tau$. Thus, the solution of this equation is, within its assumptions, a universal description of the electron energy spectrum due to beamstrahlung.

\section{Solution of the  Asymptotic Yokoya-Chen system}
\label{sec:AYC}

The AYC equations have the form of simple integro-differential equations, but it is difficult to find an exact analytical solution. Instead, we build up a semi-quantitative approximate solution, the details of which are given in Appendix \ref{sec:AYCappendix}. For the electron distribution $h_e(x, \tau)$, we find
\begin{eqnarray}
        h_e(x, \tau) =  h^{(1)}_e(x, \tau) + \bigl(1-e^{-c \tau}
          (1 + a \tau +{1\over 2}  (a\tau)^2)\bigr)\,  h_e^{(2)} (x, \tau) ,
             \label{yeformmain}
       \end{eqnarray}
with 
     \begin{eqnarray}
     h^{(1)}_e(x, \tau) = e^{-\tau} \biggl[ \delta(x-1) + {\tau\over B}  {1\over
     x^{1/3}(1-x)^{2/3}} + {B' \tau^2\over 2 B^2} {1\over
       x^{2/3}(1-x)^{1/3}} \biggr] \ ,
       \label{yonemain}
     \end{eqnarray}
       \begin{eqnarray}
      h_e^{(2)} (x, \tau) =    {b \tau \over 3 x^{2/3}} \exp[ -b \tau
         x^{1/3}]  \ ,
         \label{ytwomain}
       \end{eqnarray}
where $B' = \Gamma({1\over 3})^2/\Gamma({2\over 3}) =  5.2999$, and both $a$ and $b$ are free parameters that should be adjusted to best represent the behavior of the true solution $h_e(x, \tau)$. 

\begin{figure}
\begin{center}
\includegraphics[width=1\hsize]{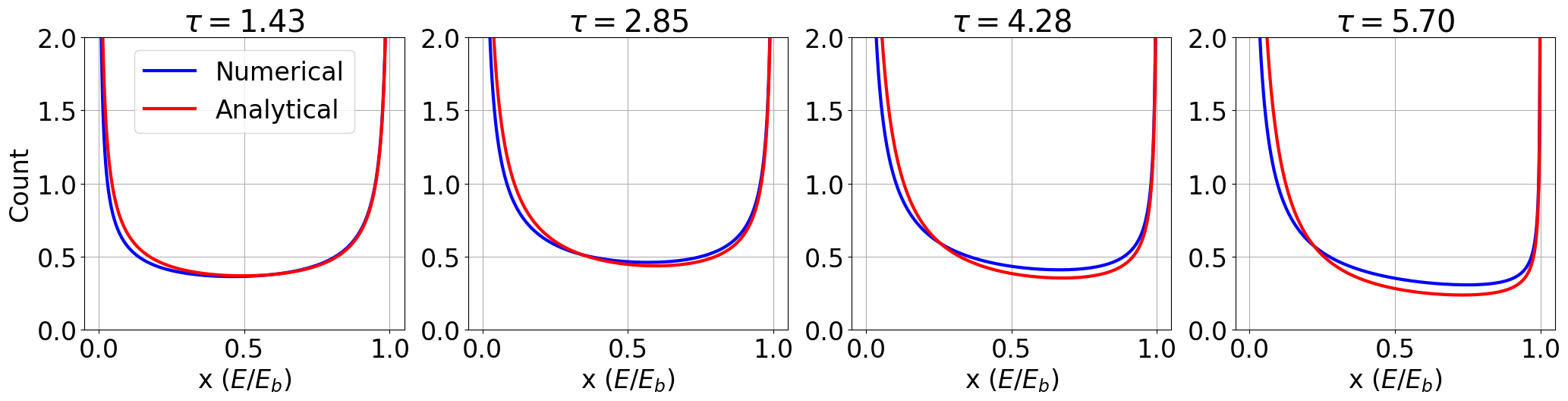}
\end{center}
\caption{Comparison of the numerical solution of the AYC equation for $h_e(x,\tau)$, in blue, to the formula \ref{yeformmain}, in red.}
\label{fig:AYCvsapprox}
\end{figure}

The expression (\ref{yeformmain}) with the parameter choices given in the appendix is compared to the numerical solution to the AYC equation in  Fig.~\ref{fig:AYCvsapprox}. The agreement is almost perfect for small $\tau$, which is expected from our derivation, and remains at least qualitatively correct over the whole range of $\tau$. 

\begin{figure}
\begin{center}
\includegraphics[width=1\hsize]{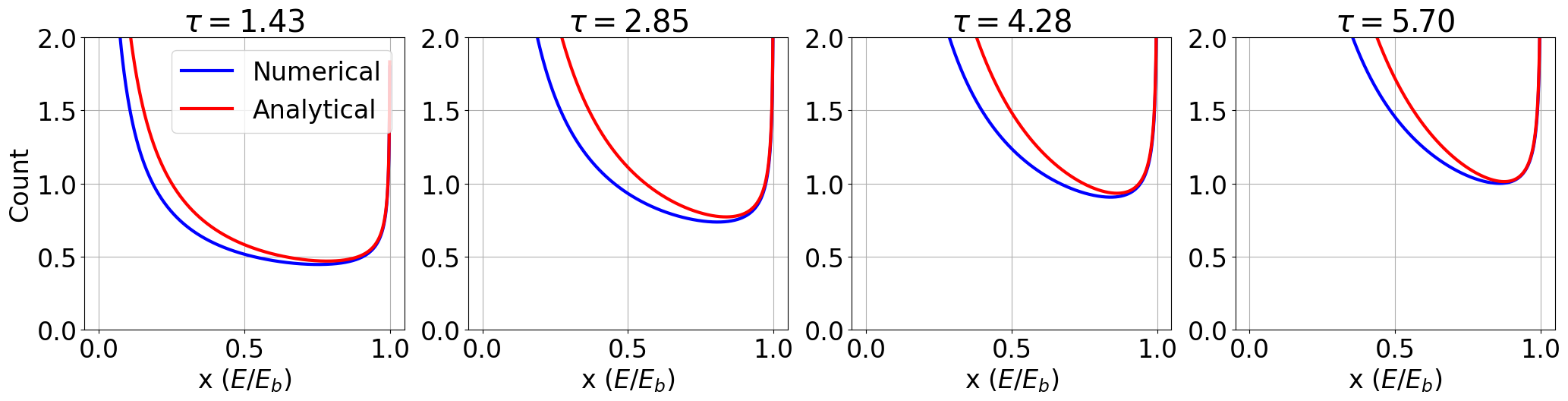}
\end{center}
\caption{Comparison of the numerical solution of the AYC equation
  for $h_g(x,\tau)$, in
  blue, to the formula \ref{yeformmain}, in red.}
\label{fig:AYCgvsapprox}
\end{figure}

We can find an approximate solution for $h_g(x,\tau)$ in a similar way. We find
\begin{eqnarray}
   h_g(x,\tau) = h_g^{(1)}(x,\tau) + h_g^{(2)}(x,\tau) +
     (1 - e^{-c\tau}(1 + c\tau + {1\over 2}(c\tau)^2)) h_g^{(3)}(x,\tau) \ ,
     \label{ygformmain}
   \end{eqnarray}
with 
   \begin{eqnarray}
     h_g^{(1)}(x,\tau) &=&   (1 - e^{-\tau}) {1\over B} {1\over
       x^{2/3}(1-x)^{1/3}} \\
h_g^{(2)}(x,\tau) &=&   (1 - e^{-\tau}(1 + \tau)) {d \over
     B} {(\log(1/x) + e) \over x^{2/3}} \nonumber   \\
   h_g^{(3)}(x,\tau) &=&    {f \log(1/x)\over x^{2/3}},
     \end{eqnarray}
and $c, d, e,$ and $f$ fit parameters given in Table~\ref{tab:fit_parameters}. This gives a reasonable representation of the solution to the AYC equation, which we demonstrate with a comparison to the numerical solution of the AYC equation in Fig.~\ref{fig:AYCgvsapprox}. 

\section{Simple AYC model for the bunch-bunch collision}
\label{sec:slices}

The solution of the AYC equation discussed in the previous section is a function of the time parameter $\tau$.   For a bunch-bunch collision, $\tau$ runs from 0 to $N_\gamma$. We then expect that $h_e(x, N_\gamma)$ should represent the energy distribution of the electron or positron bunch as it emerges from the bunch-bunch collision. However, the energy distribution of the electrons and positrons in collision events should be given by the AYC solution at an earlier time. In our model, collision events occur whenever an electron slice overlaps with a positron slice.

\begin{figure}
\begin{center}
\includegraphics[width=0.50\hsize]{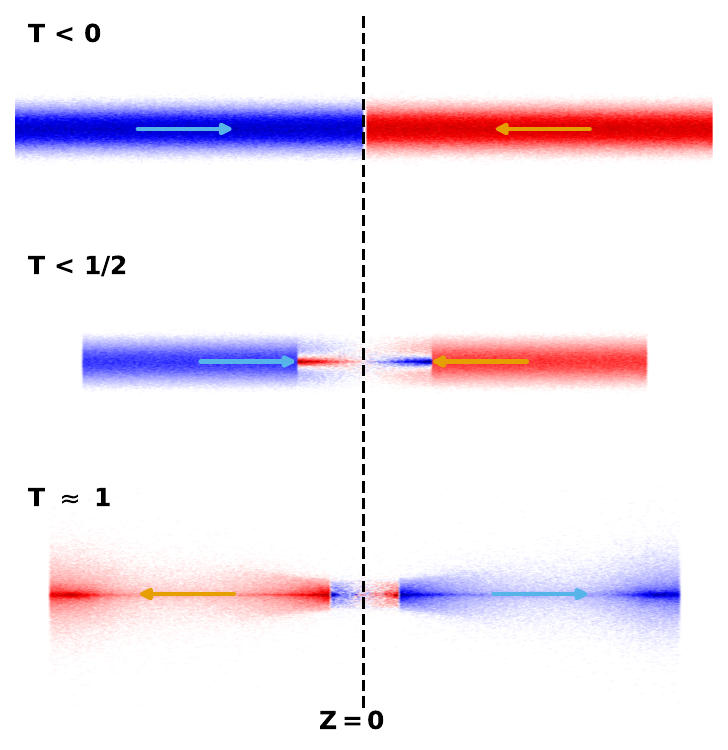}
\end{center}
\caption{Schematic drawing of the collision process in our simple model, defining the coordinates $T$ and $Z$.}
\label{fig:collision}
\end{figure}

 
As we described in Section~\ref{sec:toAYC}, we approximate the Gaussian bunches as cylinders with sharp front and back surfaces. Let the collision coordinate be $Z$, running from -1  to 1, and let $T = \tau/N_\gamma$ be an associated time variable. In these coordinates,  the collision process looks like Fig.~\ref {fig:collision}, where $Z = 0$  is the coordinate point at which the two bunches meet and the center of the collision process at later times. We continue to assume that the bunches are uniform on each transverse slice. Then, a given slice of the cylinder undergoes the Yokoya-Chen evolution as it moves through the opposite bunch. This slice is ``processed" as the fields of the opposite bunch act on it. At any intermediate time, the slice has been processed through an amount of time $T$, equal to 0 when the slice first encounters the head of the opposite bunch and equal to $1$ when it exits the tail of the opposite bunch. 

The luminosity spectrum associated with the bunch-bunch collision will then depend on integrals of the probability density of an electron with energy fraction $x_1$ meeting a positron with energy fraction $x_2$ at a given time $T$ into the collision:  $P(x_1, x_2, T)\, dx_1 dx_2 dT$. In our model, this density is given by integrals over the region where the electron and positron beams overlap, 

\begin{eqnarray}
P(x_1, x_2, T) &=&
\int_{-\bar{T}}^{+\bar{T}} dZ  \ h_e(x_1,N_\gamma(T+Z)) \
    h_e(x_2, N_\gamma (T-Z)). \label{Pfcn}
\end{eqnarray}
with $\bar{T} = T$ for $T<1/2$, and $\bar{T} = 1 - T$ for $T>1/2$. The integral of $P(x_1,x_2,T)$ over all times in the collision simplifies to the form
\begin{eqnarray}
   \int^1_0 dT \ P(x_1,x_2,T) = f_e(x_1, N_\gamma T) \cdot f_e(x_2,
   N_\gamma T),
   \end{eqnarray}
   where
   \begin{eqnarray}
     f_e(x) = \int^1_0 dT  \ h_e(x,N_\gamma T).
     \label{feval}
   \end{eqnarray}
This function can be identified with the overall electron or positron energy spectrum in collision events. The expressions for the luminosity spectra in CM energy and rapidity are then
   \begin{eqnarray}
      \frac{{\cal L}(z)}{{\cal L}_0} & =& 2z \int^1_ {z^2}  {dx_1\over x_1}  f_e(x_1)\,
      f_e\left({z^2\over x_1}\right)\nonumber   \\
     \frac{{\cal L}(y)}{{\cal L}_0} & =& 2e^{-2y}\int^1_0 dx_1\, x_1\ f_e(x_1)\,
                     f_e\left(e^{-2y} x_1 \right),
                     \label{odists}
     \end{eqnarray}
where ${\cal L}_0$ is the total luminosity. The second equation applies for $y>0$ only, and for negative values of $y$, use ${\cal L}(-|y|) = {\cal L}(|y|)$. Similar formulae also apply to the $e\gamma$ and $\gamma\gamma$ luminosity functions by replacing $f_e$ with $f_g$, the energy distribution of the photons, given by
\begin{eqnarray}
    f_g(x) = \int^1_0 dT  \ h_g(x,N_\gamma T)   \ .
    \label{fgval}
   \end{eqnarray}
   
It is important to note that because $h_e(x, T)$ is normalized to $1$, the $e^+e^-$ luminosity spectrum from our model is also normalized to $1$. On the other hand, $h_g(x, T)$ is directly calculated from the YC equation given the normalization of $h_e(x, T)$, and hence the ratio between the integrals of $h_g$ and $h_e$ reflects the true ratio between the number of photons and electrons. As a consequence, the ratio of the integrals of the $e\gamma$ and $\gamma\gamma$ luminosity spectra reflects the true ratio of $e\gamma$ and  $\gamma\gamma$ collisions to $e^+e^-$ collisions.

\begin{figure}
\begin{center}
\includegraphics[width=0.90\hsize]{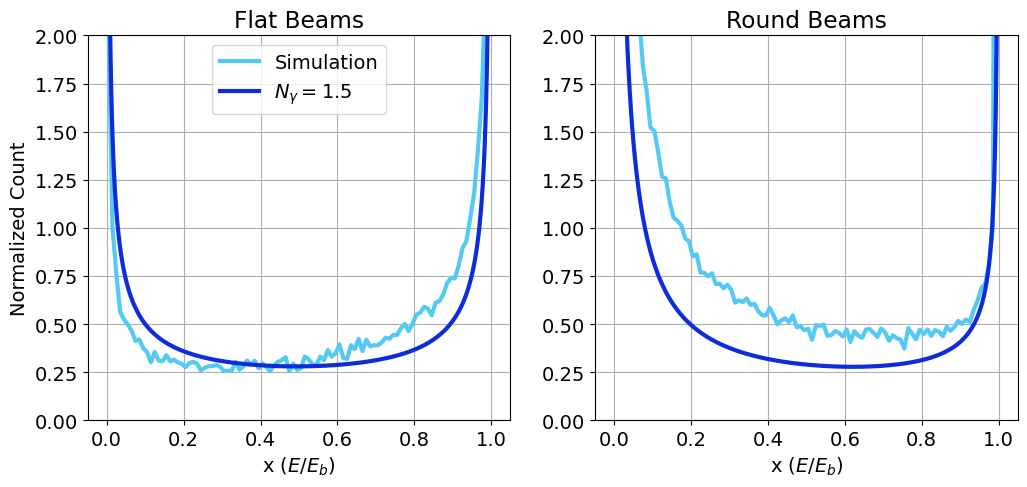}
\end{center}
\caption{Comparison of GUINEA-PIG simulation data for the electron energy distribution in $e^+e^-$ collisions with the formula \ref{feval} for $f_e(x)$, for a CM energy of 10~TeV. Left: flat beams; Right: round beams. In the analytic formulae, we use the nominal values $N_\gamma$ = 1.5 for flat beams and 5.7 for round beams.}
\label{fig:fecomp}
\end{figure}

\begin{figure}
\begin{center}
\includegraphics[width=0.90\hsize]{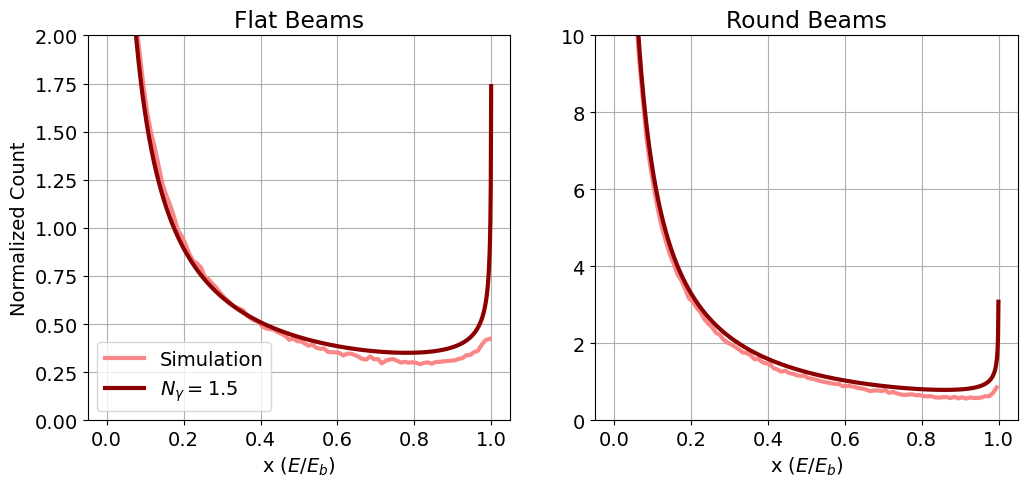}
\end{center}
\caption{Comparison of GUINEA-PIG simulation data for the photon energy distribution in $e^+e^-$ collisions with the formula \ref{fgval} for $f_g(x)$, for a CM energy of 10~TeV. Left: flat beams; Right: round beams. In the analytic formulae, we use the nominal values $N_\gamma$ = 1.5 for flat beams and 5.7 for round beams.}
\label{fig:fgcomp}
\end{figure}

\begin{figure}
\begin{center}
\includegraphics[width=0.90\hsize]{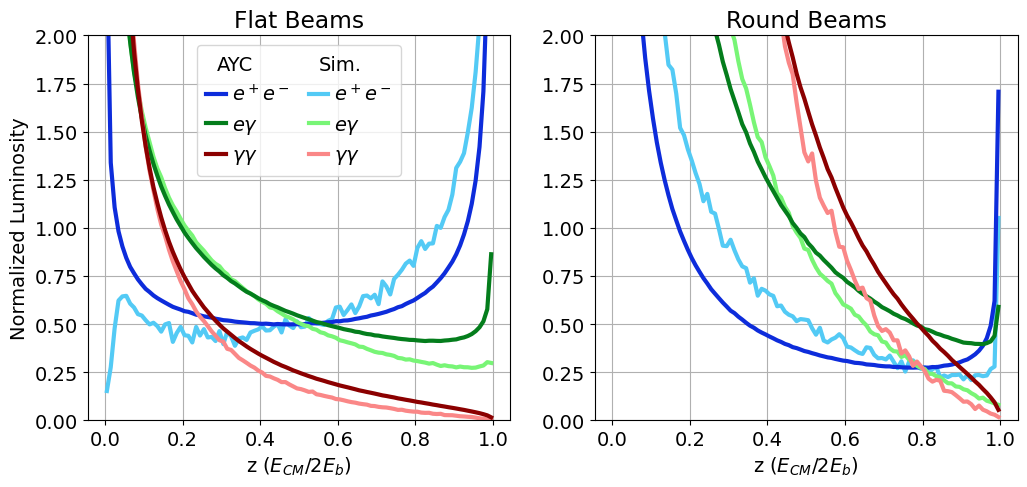}
\end{center}
\caption{Comparison of GUINEA-PIG simulation data for the CM luminosity distributions for $e^+e^-$, $e\gamma$, and $\gamma\gamma$ collisions with the predictions of the AYC model, for a CM energy of
  10~TeV: Left: flat beams; Right: round beams.}
\label{fig:zcomp}
\end{figure}

The electron and photon energy distributions in collisions predicted by the AYC model using the approximate formulae (\ref{yeformmain}) and (\ref{ygformmain}) are compared to the 10~TeV simulation data in Fig.~\ref{fig:fecomp} and Fig.~\ref{fig:fgcomp}, respectively. The predictions of the model for the luminosity spectra in $z$, compared to the simulation data, are shown in Fig.~\ref{fig:zcomp}. We see that the energy spectrum of electrons and photons in collision events is in close agreement with the simulation, given that we have simplified a complex 3-dimensional collision to a 1-dimensional equation in our model. The energy spectrum of photons disagrees slightly at high energy, and this manifests itself in the high-energy tail of the $e\gamma$ CM luminosity spectra. Otherwise, the luminosity spectra in CM energy are similar. The discrepancy in the flat beam $e^+e^-$ luminosity at low CM energy is due to a low energy cutoff in GUINEA-PIG, which does not appear in the round beam case due to the range of the plot.

Moreover, equations (\ref{odists}) predict that the expression for the luminosity factorizes into a convolution of single-beam distribution functions. This property is well satisfied by the GUINEA-PIG simulation data. The spectrum of electrons or positrons in collision events extracted from the simulation data is close to identical when considering only $e^+e^-$ or $e\gamma$ collisions, and the spectrum of photons in collision events is close to identical when considering only $e\gamma$ or $\gamma\gamma$ collisions.

\section{Time development of the bunch-bunch luminosities}
\label{sec:timeforlumi}

The model in Section~\ref{sec:slices} allows us to study the time development of the luminosity in bunch-bunch collisions. In $e^+e^-$ collisions, this allows us to understand what fraction of the luminosity is generated at the beginning of the bunch-bunch interaction, when the energy densities are strongly peaked toward $x = 1$, so that we can investigate the advantage of short bunches in the collider design~\cite{Yakimenko:2018kih}. 

\begin{figure}
\begin{center}
\includegraphics[width=0.9\hsize]{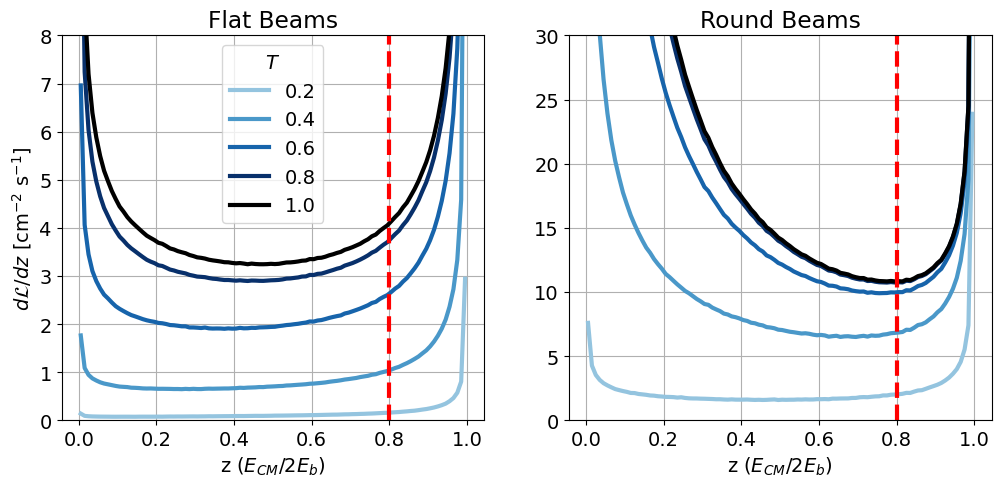}
\end{center}
\caption{Time development of the $e^+e^-$ luminosity in 10~TeV bunch-bunch collisions.
   The plots show the absolute luminosity distributions $ d{\cal L}/dz$ (in units of $10^{35}$/cm$^2$sec) integrated up to time $T = \tau/N_\gamma$, with (from the bottom)  $T = 0.2, 0.4, 0.6, 0.8, 1.0$:  Left:  flat beams; Right: round beams.  The vertical red line marks the $z > 0.8$  (${\cal L}_{20}$) region.}
\label{fig:eelumiint}
\end{figure}

The time evolution of the $e^+e^-$ luminosity spectrum from our model can be obtained by integrating the probability density function $P(x_1,x_2,T)$ in equation (\ref{Pfcn}) up to intermediate values of $T$ between 0 and 1. The results for the luminosity spectrum in $z$ are shown in Fig.~\ref{fig:eelumiint}. These plots are normalized such that the $\mathcal{L}_{20}$ at the final time matches the $\mathcal{L}_{20}$ from the simulation data, as given in Table~\ref{tab:outparams}. 

There is a key difference between the flat beam and the round beam case. In the flat beam cases, the $e^+e^-$ luminosity for all values of $z$ is generated fairly uniformly through the bunch-bunch collision process.  In the round beam case, the luminosity at high $z$ is generated almost completely in the first 60\% of the bunch-bunch collision. The luminosity generated in the
latter part of the collision is almost all at low $z$ and contributes mainly to the beam-related backgrounds. Note that since the generation of beamstrahlung depends only on the parameter $N_\gamma$, this effect only depends on $N_\gamma$, with uniform generation of the luminosity being a feature of designs with $N_\gamma \lesssim 3$.

We can investigate whether it is possible to mitigate beamstrahlung by decreasing the bunch length~\cite{Yakimenko:2018kih}. With the understanding that beamstrahlung effects depend only on $N_\gamma$, this reduces to the scaling of $N_\gamma$ with $\sigma_z$. Let $a = \sigma_z/\sigma_{z0}$ be the reduction of the bunch length. This increases the particle density in each bunch by a factor $a$. Then, $N_\gamma$ scales as
\begin{eqnarray}
         \nu_{cl}  \Upsilon^{-1/3} \sigma_z \sim   a\cdot
         a^{1/3} \cdot a^{-1}   \sim    a^{1/3}  \ .
\end{eqnarray}
Hence, by changing the bunch length, we obtain only a slight advantage in the reduction of beamstrahlung.


\begin{figure}
\begin{center}
\includegraphics[width=0.90\hsize]{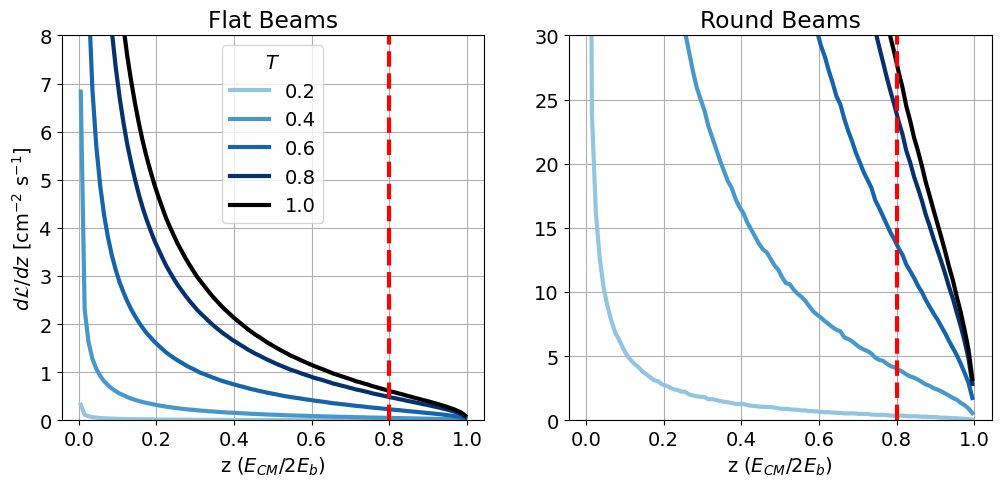} \ \

\end{center}
\caption{Time development of the $\gamma\gamma$ luminosity in 10~TeV bunch-bunch collisions.
   The plots show the absolute luminosity distributions $ d{\cal L}/dz$ (in units of $10^{35}$/cm$^2$sec) integrated up to time $T = \tau/N_\gamma$, with (from the bottom)  $T = 0.2, 0.4, 0.6, 0.8, 1.0$:  Left:  flat beams; Right: round beams.   The vertical red line marks the $z > 0.8$  (${\cal L}_{20}$) region. }
\label{fig:gglumiint}
\end{figure}

\begin{figure}
\begin{center}
\includegraphics[width=0.9\hsize]{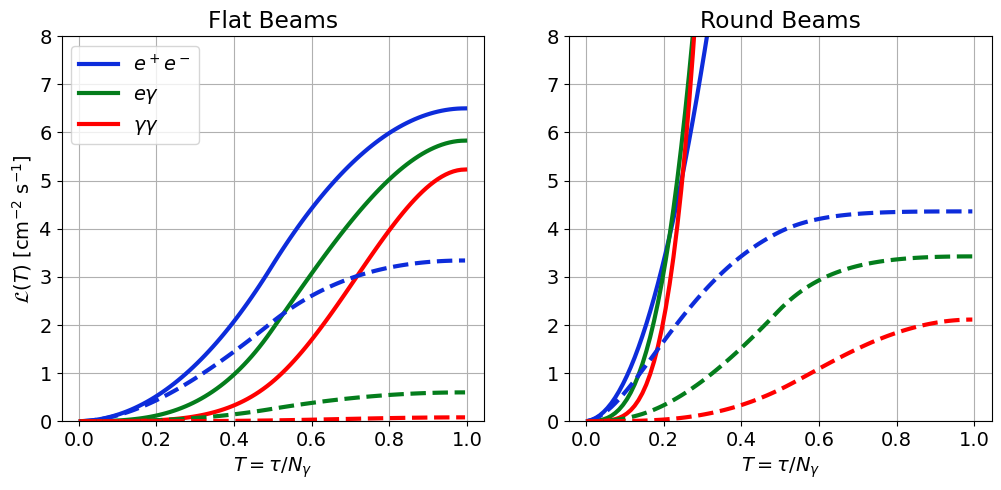} 
\end{center}
\caption{Accumulated luminosity ${\cal L}(T)$, in units of $10^{35}$/cm$^2$sec as a function of $T = \tau/N_\gamma$.  The luminosities for $e^+e^-$, $e^-\gamma$, and $\gamma\gamma$ collisions are shown with blue, green, and red curves, respectively.  The solid lines shows the total luminosity, while the dashed lines show the ${\cal L}_{20}$.   Left:  flat beams; Right:  round beams.}
\label{fig:lumiwT}
\end{figure}

In Fig.~\ref{fig:gglumiint}, we show the development of the $\gamma\gamma$ luminosity in absolute units at values of $T$ between 0 and 1. This luminosity is larger in the round beam case, but the contribution at high $z$ always remains a small fraction of the $e^+e^-$ luminosity. This demonstrates that designing an $e^+e^-$ collider with substantial $\gamma\gamma$ luminosity at the highest energies will be a challenge.

The time-dependence of the total and ${\cal L}_{20}$ luminosity is shown in Fig.~\ref{fig:lumiwT}. It follows our intuition that the $e^+e^-$ luminosity turns on first, followed by the $e\gamma$ and then the $\gamma\gamma$ luminosity as the $\gamma$  distribution builds up during the collision. Note that in the round beam case, the $\mathcal{L}_{20}$ for all three collision types reaches an asymptote before the end of the collision, which is desirable. After a certain time in the collision, it is no longer possible to collide particles with high CM energy, as all the particles lose energy throughout the collision. 


\section{Conclusions}
\label{sec:conclusions}

In this paper, we utilize an analytical model of beamstrahlung for high $\Upsilon$, inspired by data from particle-in-cell simulations of bunch-bunch collisions, to simplify the description of beamstrahlung radiation and the associated effects on the collider luminosity spectra. We have shown that, for the large values of $\Upsilon$ that are realized at multi-TeV $e^+e^-$ colliders, the spectrum of beamstrahlung radiation is determined by the universal functions $h_e(x,t)$ and $h_g(x,t)$, with the total amount of radiation depending only on the parameter $N_\gamma$.

As we noted in the introduction, the understanding of beamstrahlung is only the first step in the study of the bunch-bunch collision at high energies. With our understanding of beamstrahlung emissions as the backbone, we must add incoherent bremsstrahlung, coherent and incoherent $e^+e^-$ pair production, and other relevant QED processes. In addition, we note that our picture of beamstrahlung, used both in the GUINEA-PIG simulations and in our analytic theory, is based on a simplified understanding of photon emission in the bunch-bunch collision, using the Local Constant Field Approximation (LCFA) \cite{Chen:1987,Piazza_LCFA2, Piazza_LCFA} and the assumption that each photon emission can be treated independently. We may seek to reexamine this assumption in the future.

\begin{figure}
\begin{center}
\includegraphics[width=0.90\hsize]{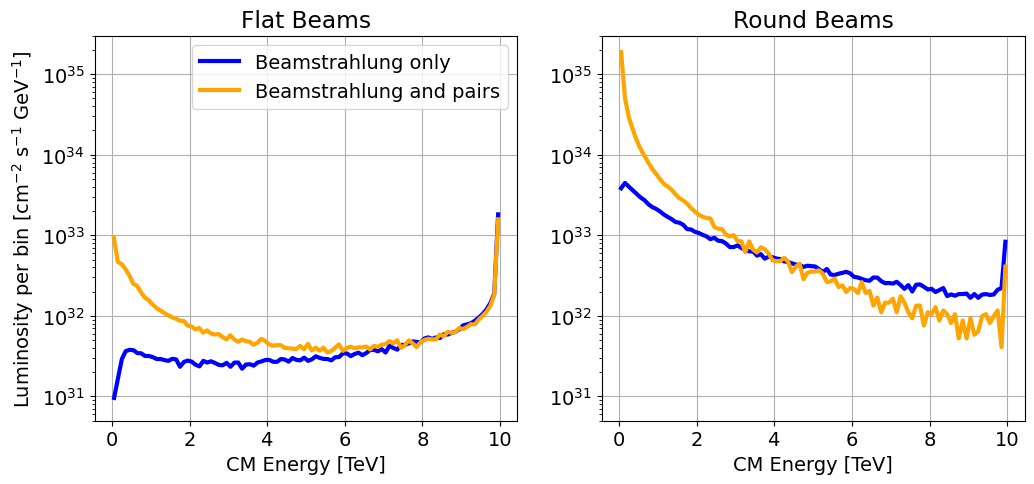}
\end{center}
\caption{Comparison of the luminosity spectra from the 10~TeV
  GUINEA-PIG simulations showing the effect of  including
  pair-production
  processes: blue curves: beamstrahlung only; orange curves:
  including pair production.}
\label{fig:turnonpairs}
\end{figure}

The effect of $e^+e^-$ pair production can be significant, even for the luminosity function at high $z$.  This is illustrated in Fig.~\ref{fig:turnonpairs}, which shows the effect on the luminosity distribution in $z$ for the flat and round beam 10~TeV simulation when one turns on pair production.  In future work, we will explore the effects of these additional QED processes~\cite{Esberg:2014}, based on the foundation that the theory in this paper provides.

Moreover, in some of the simulations in this paper, the value of $\Upsilon$ is well within the strong field QED regime, characterized by $\alpha\Upsilon^{2/3} \gtrsim 1$ \cite{Fedotov:2017, Mironov_SFQED}. In this regime, the formulas used for quantum beamstrahlung based on perturbative QED are no longer valid, and one must apply the strong-field QED theory or a calculation based on lattice QED \cite{Kogut:2024}. We recognize this as a limitation of this work, and in future works we will explore the non-perturbative regime based on the theory in this paper. 

Finally, we are currently developing a high-performence GPU-based beam-beam tool within the particle-in-cell code WarpX \cite{vay_2025_17539393}, which will be used to perform detailed modeling of future energy-frontier $e^+e^-$ colliders.

\begin{acknowledgements}

The authors would like to thank Sebastien Meuren, Thomas Grismayer, Jens Osterhoff, and Remi Lehe for useful conversations. DH would like to thank the U.S. Department of Energy's SULI program for the opportunity to be involved in this research. The work of DH, SG, and MEP is supported by the U.S. Department of Energy under contract DE--AC02--76SF00515. 

This material is based upon work supported by the CAMPA collaboration, a project of the U.S. Department of Energy, Office of Science, Office of Advanced Scientific Computing Research and Office of High Energy Physics, Scientific Discovery through Advanced Computing (SciDAC) program. This research used resources of the National Energy Research Scientific Computing Center (NERSC), a Department of Energy Office of Science User Facility using NERSC award HEP-ERCAP0031191.

\end{acknowledgements}

\appendix

\setcounter{table}{0}
\renewcommand{\thetable}{A\arabic{table}}

\section{Solution of the Asymptotic Yokoya-Chen system}
\label{sec:AYCappendix}
  
In Section \ref{sec:toAYC} we arrive at a universal differential equation for the evolution of the electron and photon energy spectrum. Here we give the details of the approximate analytical solution to these differential equations presented in Section \ref{sec:AYC}.
   
Starting from the initial condition
   \begin{eqnarray}
   h_e(x, \tau = 0) = \delta(x-1),
   \end{eqnarray}
we develop a solution for early times. For $\tau\ll 1$, at first-order, the solution is
   \begin{eqnarray}
   h_e(x,\tau) =  (1 - \tau)\delta(x-1) + {\tau\over B} {1\over
     x^{1/3}(1-x)^{2/3}}.
     \end{eqnarray}
To continue this process, we need the integrals
     \begin{eqnarray}
     \int^1_x {d\bar x\over x^{1/3} (\bar x - x)^{2/3} \bar
       x^{1/3}} \cdot  {1\over
     \bar x^{1/3}(1-\bar x)^{2/3}} &=& B' {1\over x^{2/3}
       (1-x)^{1/3}}\nonumber   \\
     \int^1_x {d\bar x\over x^{1/3} (\bar x - x)^{2/3} \bar
       x^{1/3}} \cdot  {1\over
     \bar x^{2/3}(1-\bar x)^{1/3}} &=& B {1\over x},
     \end{eqnarray}
where $B' = \Gamma({1\over 3})^2/\Gamma({2\over 3}) =  5.2999$. After two iterations, the singularity at $x=1$ has disappeared; all further integrals are regular as $x\to 1$. Also, notice that the second integral does not actually generate a $1/x$ singularity as $x\to 0$, since the generated term would be
     \begin{eqnarray}
     \left( -1 + {B\over B}\right){1\over x} = 0 \ .
     \end{eqnarray}

For the terms singular as $x \to 1$, we can approximate the first term on the right-hand side by replacing $-h_e/x^{1/3}$ by $-h_e$.  Then we would solve the equation
       \begin{eqnarray}
       {\partial\over \partial \tau} h_e(x,\tau)  =  -  h_e(x, \tau)
                 +  {1\over B} \int_x^1 d\bar x\
                   {1\over  x^{1/3} (\bar x - x)^{2/3} \bar x^{1/3}}
         \    h_e(\bar x, \tau) \ ,
         \label{aaYC}
                   \end{eqnarray}
We can look for a solution of the form
                   \begin{eqnarray}
              h_e(x, \tau) &=&  A(\tau) \delta(x-1) + B(\tau) {1\over
                x^{1/3} (1 - x)^{2/3} }  \nonumber   \\
              & & \hskip 0.4in  + C(\tau){1 \over  x^{2/3} (\bar x -
              x)^{1/3}} + \mbox{regular\ as}\ x\to 1
            \end{eqnarray}
The solution is
     \begin{eqnarray}
     h^{(1)}_e(x, \tau) = e^{-\tau} \biggl[ \delta(x-1) + {\tau\over B}  {1\over
     x^{1/3}(1-x)^{2/3}} + {B' \tau^2\over 2 B^2} {1\over
       x^{2/3}(1-x)^{1/3}} \biggr] \ ,
       \label{yone}
     \end{eqnarray}
plus terms regular as $x\to 1$. This function contains all terms of the solution to the  AYC equation that are singular as $x\to 1$. 
 
On the other hand, the iteration shows that $h_e(x,T)$ is a Laurent series in the variable $x^{1/3}$ with highest power $x^{2/3}$ as $x\to 0$.  A simple function with the qualitatively correct behavior is
       \begin{eqnarray}
      h_e^{(2)} (x, \tau) =    {b \tau \over 3 x^{2/3}} \exp[ -b \tau
         x^{1/3}]  \ ,
         \label{ytwo}
       \end{eqnarray}
with $a$ a free parameter.  This satisfies
       \begin{eqnarray}
       \int_0^\infty dx \ h_e^{(2)} (x, \tau) = 1 \ ,
       \end{eqnarray}
and so is approximately normalized on the interval  $[0,1]$ in $x$.

The solution transitions from $h_e^{(1)}(x, \tau)$ to $h_e^{(2)}(x, \tau)$ at intermediate values of $\tau$. The term $h_e^{(2)}(x, \tau)$ turns on as $\tau^2$.  Then the solution of the AYC equation should approximately have the form
        \begin{eqnarray}
        h_e(x, \tau) =  h^{(1)}_e(x, \tau) + \bigl(1-e^{-c \tau}
          (1 + a \tau +{1\over 2}  (a\tau)^2)\bigr)\,  h_e^{(2)} (x, \tau) ,
             \label{yeform}
       \end{eqnarray}
where both $a$ and $b$ are free parameters that should be adjusted to best represent the behavior of the true solution $h_e(x, \tau)$. These values are given in Table~\ref{tab:fit_parameters}. The expression (\ref{yeform}) is an elementary and easy to integrate function that can be readily applied in the further analysis of this paper. Note that it does not exactly satisfy the normalization in (\ref{ynorm}); the discrepancy is about 5\% at late times.

For $h_g(x,\tau)$ we insert the delta function term in (\ref{yone}) into the right-hand side of (\ref{AYCg}), giving the photon distribution
   \begin{eqnarray}
     h_g^{(1)}(x,\tau) =   (1 - e^{-\tau}) {1\over B} {1\over
       x^{2/3}(1-x)^{1/3}}.
     \end{eqnarray}
This is the only term in $h_g(x,\tau)$ that is singular as $x\to 1$.  The integrals of the second and third terms in \ref{yone} are not simple, but they suggest a functional form for the remaining terms.  We find that, with
\begin{eqnarray}
   h_g^{(2)}(x,\tau) &=&   (1 - e^{-\tau}(1 + \tau)) {d \over
     B} {(\log(1/x) + e) \over x^{2/3}} \nonumber   \\
   h_g^{(3)}(x,\tau) &=&    {f \log(1/x)\over x^{2/3}},
   \end{eqnarray}
the sum
   \begin{eqnarray}
   h_g(x,\tau) = h_g^{(1)}(x,\tau) + h_g^{(2)}(x,\tau) +
     (1 - e^{-c\tau}(1 + c\tau + {1\over 2}(c\tau)^2)) h_g^{(3)}(x,\tau) \ ,
     \label{ygform}
   \end{eqnarray}
with the fit parameters given in Table~\ref{tab:fit_parameters} give a reasonable representation of the function.

\begin{table}[b]
\caption{\label{tab:fit_parameters}Fit parameters extracted from the model.}
\begin{ruledtabular}
\begin{tabular}{l c}
\hline
\textbf{Parameter} & \textbf{Value} \\
\hline
$a$ & 1.0 \\
$b$ & 0.7 \\
$c$ & 0.5 \\
$d$ & 0.4 \\
$e$ & 3.0 \\
$f$ & 0.7 \\
\hline
\end{tabular}
\end{ruledtabular}

\end{table}

\section{Analytic formula for the electron and photon distribution
  functions}

In (\ref{yeform}) and (\ref{ygform}), we gave analytic parameterizations of the solutions of the AYC equations for electrons and photons. We found
  \begin{eqnarray} 
  h_e(x, \tau) &=&  e^{-\tau} \biggl[ \delta(x-1) + {\tau\over B}  {1\over
     x^{1/3}(1-x)^{2/3}} + {B' \tau^2\over 2 B^2} {1\over
       x^{2/3}(1-x)^{1/3}} \biggr] \nonumber   \\
       & & \hskip 0.1in + \bigl(1-e^{-a \tau}
        (1 + a \tau +{1\over 2} (a\tau)^2)\bigr)\,    {b \tau \over 3 x^{2/3}} \exp[ -b \tau x^{1/3}]  \nonumber   \\
h_g(x, \tau) &=&  (1 - e^{-\tau}) {1\over B} {1\over
       x^{2/3}(1-x)^{1/3}} +  (1 - e^{-\tau}(1 + \tau)) {d \over
     B} {(\log(1/x) + e) \over x^{2/3}}\nonumber   \\
  & & \hskip 0.1in +       
 (1 - e^{-c\tau}(1 + c\tau + {1\over 2} (c\tau)^2))
      {f \log(1/x)\over x^{2/3}}
      \label{yform}
       \end{eqnarray}  
where   $ B = 2\pi/\sqrt{3} = 3.6276$, $B' = \Gamma[{1\over 3}]^2/\Gamma[{2\over 3}] = 5.2999$.
 
It is straightforward to integrate the functions $h_e(x,t)$ and $h_g(x,t)$ over time to compute the effective parton distribution functions $f_e(x)$, $f_g(x)$ as functions of $N_\gamma$. It is easiest to give the solutions in terms of the functions
\begin{eqnarray}
    e_0(T) &=& \int_0^T dt \  e^{-t}  =   1 - e^{-T} \nonumber   \\
   e_1(T) &=& \int_0^T dt \ t \, e^{-t}  =   1 - e^{-T}(1 + T)\nonumber   \\
   e_2(T) &=& \int_0^T dt \ (t^2/2)\,  e^{-t}  =   1 - e^{-T} (1 + T +
   T^2/2)  \nonumber   \\
   e_3(T) &=& \int_0^T dt \ (t^3/6)\,  e^{-t}  =   1 - e^{-T} (1 + T +
   T^2/2 + T^3/6) \ . 
  \end{eqnarray}
Then,
   \begin{eqnarray}
   f_e(x,N_\gamma) &=&  {1\over N_\gamma}\biggl[  e_0(N_\gamma) \
   \delta(x-1) + e_1(N_\gamma) {1\over B x^{1/3} (1-x)^{2/3}} + e_2(N_\gamma)
                  {B'\over B^2 x^{2/3} (1-x)^{1/3}} \nonumber   \\
                  & & \hskip 0.1in + e_1(b x^{1/3} N_\gamma){1\over (3 b x^{4/3})}
                  -\biggl\{ e_1((a + b x^{1/3}) N_\gamma) {b\over
                      3x^{2/3}(a + bx^{1/3})^2}
                                 \nonumber   \\
       & & \hskip 0.1in +  e_2((a + b x^{1/3}) N_\gamma){2 ab\over 
       3x^{2/3}(a + bx^{1/3})^3 }
       + e_3((a + b x^{1/3}) N_\gamma) { a^2b\over x^{2/3}(a + bx^{1/3})^4 } \biggr\} \biggl] \ ,
         \nonumber   \\
         \end{eqnarray}
and
           \begin{eqnarray}
   f_ g(x,N_\gamma) &=&  {1\over N_\gamma}\biggl[ ( N_\gamma - e_0(N_\gamma))
  {1\over B x^{2/3} (1-x)^{1/3}} \nonumber   \\
                  & & \hskip 0.1in +  (N_\gamma - (e_0(N_\gamma) +
                  e_1(N_\gamma)))\ 
                  {d\over B}  \ {(\log(1/x) + e )\over x^{2/3}} \nonumber   \\
               & & \hskip 0.1in   + (N_\gamma  - {1\over c}(e_0(c N_\gamma) + e_1(c N_\gamma) + e_2(c N_\gamma))) 
               {f \log(1/x)\over x^{2/3}} ) \biggr] \  .
                  \end{eqnarray}

\section{Simulation details}

The parametric scan over the aspect ratio has been performed with $n_m = 10^7$ macroparticles in a domain of $ 9 \sigma_x \times 30 \sigma_y \times 3.5 \sigma_z $, with a transverse resolution of $n_x = 180,  n_y = 600$ and $n_z = 210$ $z$-slices in GUINEA-PIG~\cite{Schulte:1999tx}. The number of substeps is $n_t = 1$. The simulation inputs and results can be found on Zenodo~\cite{peskin_2025_17716686}.


\bibliographystyle{apsrev}
\bibliography{bibliography}

\end{document}